# The History, Current Status, Benefits, and Challenges of 3D Printed Organs

Alicia Shin

1. *Torrey Pines High School, 3710 Del Mar Heights Rd, San Diego, CA 92130, United States*

alicia.hs1110@gmail.com

## Abstract

There is an increase in demand for organs as transplantation is becoming a common practice to elongate human life. To reach this demand, three-dimensional bioprinting is developing from prior knowledge of scaffolds, growth factors, etc. This review paper aims to determine the current status and future possibilities of three-dimensional bioprinting of organs and evaluate the benefits and challenges, along with the history of its development. Prior research has viewed three-dimensional bioprinting as a technology that will enable safer transplantation without graft rejection and provide demand-based production. However, it faces challenges such as the need to improve biocompatibility and biofunctionality, legal and ethical issues, and the need to improve the technology itself. While the development of three-dimensional printing organs is not yet completed, we are seeing improvements and expecting it to be clinically applied soon.

## Introduction

A new concept of artificially engineered organs stands in the medical spotlight today, as it might open a new paradigm for lengthening human life. Organs are crucial parts in maintaining homeostasis within the human body; they are created when homogeneous cells develop into tissues and when these unique tissues develop into organs[17]. Unfortunately, organs are easily damaged, causing various health concerns despite going through thousands of years of evolution.

Tissue engineering (TE) has been considered as the solution to the drawbacks of organ transplantation in several ways. With the involvement of cell sources, scaffolds, and the right growth factors, tissue engineering generates customized tissues and body structures, prolonging human lives[13]. It can also contribute to the 50,000 waitlisted patients on the US transplant waiting list and decrease the annual 730,000 deaths caused by organ diseases[1]. Unlike traditional organ transplantation which uses donated organs, three-dimensional organ printing uses engineered tissues to substitute the damaged area, allowing safer and more efficient regeneration of the damaged cell. Also, organ transplantation faces a high chance of graft rejection, as well as surgical reconstruction that leads to long-term complaints. However, as tissue engineering uses patient's cells to form specialized tissues, it offers no graft rejection and healthier results.

As mentioned above, organ failure is a major contributor to worldwide mortality today[17]. While medical advancements in numerous fields have been made over past centuries, organ failure is still struggling due to factors such as organ shortage, graft rejection, and ethical conflicts[4,6,17]. However, three-dimensional printing is now one of the most popular technologies which are being developed to produce artificial organs–engineered devices that can substitute original organs through transplantation[16]. The technology attempts to mimic the natural cellular architecture by printing bio-fabricated organs on a digital model and minimizing immune responses through personalization[1]. These man-made organs will assist the recipients and their damaged organs to regain function and shape[13].

Although this technology may solve major problems including health issues and organ shortages, it has received relatively little attention in the past due to unclear visions of its future status. Currently, artificial organs show great commercial profit, and research has been steadily increasing in number, indicating that artificially manufactured organs are becoming more popularized, soon hoping to develop as a major field in medicine[11,17].

*Hypothesis and Aim*

This research reviews the current status, future insight, and limitations of three-dimensional organ printing.

## History of Organ Engineering

Tissue Engineering (TE) has been a new concept since the 1980s when the earliest clinical application occurred, forming skin tissue using fibroblasts, keratinocytes and human cells.[6] This gained more attention in the late 1990s, as 'regenerative medicine' came into place when two research groups developed human embryonic stem cells and embryonic germ cell lines.[13] Its main aim was to repair real biological tissues through cells, growth factors, and other materials.[7,16]

Scaffolds were used for cell expansion and differentiation, the process which cells would separate in the cell lineage and form diverse parts of the tissue.[13] As cells develop into artificial tissues and organs, the scaffolds assist biosynthesis of cells where enzymatic reaction produced natural products necessary for human survival.[6] Scaffolds mimic the extracellular matrix (ECM) which provides structural and biochemical support to surrounding cells. This guides tissues to develop with the right geometrical and architectural shape during regeneration. In addition, with the discovery of stem cells and technological advancements, the development of tissue engineering has constantly continued, where three-dimensional printing is now opening a new pathway to artificially engineered organs.

Although several trails adapted artificial organs, most included artificial hearts, pacemakers, dialysis for kidneys, and prosthesis. In 1996, a patient received an organ that was generated with only biomaterials for the first time. As scaffolds, a main component in generating artificial organs, faced excessive demands and constant threats of infection due to immunogenicity, interest in three-dimensional printing grew.[13] Furthermore, in 1999, the first three-dimensional-printed bladder (made with patient's cells) was transplanted, provoking a scientific revolution as it promised no graft rejection. Researchers also realized stem cells' potential to become specialized cells that can then be manipulated to treat many diseases. This has expanded to where researchers are now attempting to build original organs. As a result, in 2016, the first modern three-dimensional bioprinter was developed, specifically designed to print living tissue structures. It is known that three-dimensional printing creates a more elaborate model for organ repair, restoration, and regeneration; while it is still in development, it is receiving worldwide attention in the hopes of providing a safer and more affordable alternative for a failing organ.[16]

*Methods of 3D Organ Printing*

There are several methods researchers use to approach three-dimensional organ printing. There are two types of cells used as the building block in 3D printing: autologous and allogenic/xenogeneic cells. Autologous cells are harvested from patients themselves, offering a more appropriate and safer option for organ printing.[6] On the other hand, allogenic/xenogeneic cells are cells from different people or animals and have a high potential for graft rejection.[6]

The inkjet method is divided into two main processes. Continuous inkjet printing is accomplished by applying pressure to the bioink and adding an electric field onto the substrate to form the polymer construct.[1] This method is efficient as it pays less attention to specific forms, and the materials that weren't used to form artificial organs can be reused for later purposes. The second inkjet method involves the drop-on-demand fashion, which is very similar to continuous inkjet printing, except that it prints in a pulsed nature.[1] Due to this, the drop-on-demand method is more suitable for bioprinting.

The extrusion-based 3D bioprinting is a process in which the apparatus extrudes materials from a heated nozzle and generates a 3D structure layer by layer.[1] Extrusion is more widely known as Fused Deposition Modeling which was registered by Stratasys Ltd. This technology ensures that the material and the nozzle move uniformly and that they are constantly changing.

Lastly, laser-assisted 3D bioprinting is usually used to deposit bioink including cells onto a substrate for growth.[1] This technology ensures that the materials are deposited without contact, and usually involves a pulsed laser source, the biological material that needs to be deposited, and the substrate which receives the materials.[1]

*Current Status of Organ Printing*

Currently, three-dimensional bioprinting is already used to create simple tissues for research that supports simple regenerative medicine. In fact, Yuki Kanno's team created customized artificial bones through 3D printing and applied them to 20 patients who maintained good conditions.[14]

However, there is still a long way to go for these technologies to create complete and defined organs that can mimic the original organs' functions. Three-dimensional bioprinting will encounter challenges due to a lack of knowledge and previous cases for researchers to follow.

## Benefits of Tissue Engineering

Mainly, three-dimensional-bioprinted organs show better therapeutic effects as it enables faster healing and fewer complications after transplant.[14] Custom-made organs developed with the patient's cells provide several advantages: it allows for faster patient recovery by offering no graft rejection and allows for a reduction in surgery preparation time.[6,15] Graft rejection is a reaction when the body's natural immune system attacks a transplanted organ because it does not match the recipient's body. Unfortunately, repeated acute rejection leads to extensive scars on the tissues and overall organ, and there is no cure for this rejection.[9]

To minimize the impact of graft rejection, donated organs require immunosuppressive therapy to decrease the body's immune response, and often involves strong drugs that cause stress on other parts of the patient's body as well.[12] The initial induction phase involves high dosage while the later maintenance phase uses drugs in the long term but with less dosage.[3] Although the combination of drugs and the dosage varies on the patient's conditions, because this drug is non-specific, it reduces the entire immune system's ability, increasing the chances of lethal infection. Beneficially, three-dimensional-bioprinted organs completely dismiss these steps, making them an ideal alternative for the current donor organ transplantation.

Three-dimensional bioprinting also enables prescription of personalized medicine, including personalized drugs that considers individual genetic analyzation, which can thereby increase the effects of the drug and to faster recovery with fewer complications.[8] It also allows researchers to create personalized drug delivery, enabling the medical field to treat more patients efficiently. Furthermore, three-dimensional bioprinting promises the possibility of replacing an individual's bones, cartilages, blood vessels, and internal organs, with significantly low complications without the risk of an immune response.[8]

Lastly, researchers can tackle challenges that include vascularization of tissues, gas and nutrient exchange, biocompatibility and biodegradability of the material that is used as the substrate, shape fidelity, and preservation of functionality of the printed tissue.[1] With these improvements, the manufacturing costs can decrease, and make organ transplantations more accessible to the public.[15]

Another main benefit of perfecting three-dimensional bioprinting is solving the organ shortage crisis. 18 people die each day while waiting for a donor.[6] In addition, according to the Health Resources and Services Administration (HRSA), a new patient is added to the donor organ waiting list every nine minutes. With artificially engineered organs, ending the organ shortage crisis seems possible since the medical field can produce organs based on demand. This extends to the issue of organ trafficking: the practice of trafficking human beings for organ removal.[10] The harvested organs are usually sold on the black market for profit, and thousands of people, especially those who are not exposed to adequate protection, are at the risk of being the victim of this crime. In fact, about $1.5 billion was made in profit from 12,000 illegal transplants.[10] These statistics reveal that organ shortage is a major issue and that if three-dimensional bioprinting is popularized, both risks for patients and donors may decrease.

The main advantages of three-dimensional bioprinting listed above show why it gains support from a diverse field of science.

## Challenges of Tissue Engineering

Three-dimensional bioprinting contains several drawbacks, however. A main concern is the inability to distinguish the use of technology between therapy and human improvement.[8] There have been previous examples of technology used in different ways than its intended purpose, causing more harm than good. The same circumstances apply to three-dimensional bioprinting: it may contain positive intentions but can be taken advantage of easily. For example, technological immortality branches into two distinct ideas: rejuvenation technology and consciousness transfer technology. Rejuvenation technology attempts to prevent humans from aging, while consciousness transfer technology tries to remove the human personality using digital models. Regenerating organs typically fall in rejuvenation technology as it assists the human desire to live longer and can cause concerns due to this aspiration.

Technological challenges also exist; although 3D printed artificial organs have made advances through biomedical engineering, it still needs better biocompatibility and biofunctionality.[6] Biocompatibility is how the materials are compatible with living tissues, and biofunctionality is the ability to have a biological function in the human body. Although there are successful trials and small implications today, the big application of 3D printed organs does not promise these two important criteria. Therefore, researchers are facing challenges in finding materials that are both compatible and functional when processed into an organ.

In addition, there are ethical and legal issues in preserving personal genetic information, specifically in collecting the information and storing it.[8] As three-dimensional printing involves several experts, it is likely for a patient's information to be shared, and without proper care, leakage is inevitable. Moreover, patients involved in three-dimensional bioprinting trials may face challenges when trying to withdraw.[8] If they already had an engineered organ implanted, it is nearly impossible to undo the transplantation or replace it.[8] To improve these circumstances, more attention and development are needed, but the current inadequate tissue preservation system restricts further research and investigation. Despite the challenges, the industry is valued at $1.82 billion as of 2022.[5]

## Conclusion

The future of tissue engineering, especially the three-dimensional printing of organs, remains ambiguous. While the technology itself sounds plausible, real-life applications have been slowly progressed, and it faces several challenges listed above. The most crucial part of three-dimensional printing is the need to be further stabilized on using calcium ions during long-term in vitro cultures.[17] It also needs additional research to ensure that it can be clinically approachable.

We have previously seen how this field of technology will bring together modern science and technology, opening a diversity of advantages to humans.[16] Most importantly, it will enable patients to receive organ transplantations safe from any types of rejection, leading to increased successful surgeries. It can also enable personalized presurgical and postsurgical tools for everyone's unique health cases.[2] While it drastically increases those who receive risk-free transplantation, three-dimensional bioprinting of organs can also lead to new research methods. Moreover, there is a high possibility for research with the emergence of bio-ink materials that provide effective designs of scaffolds to be used in the extracellular matrix.[14] Researchers can try clinical experiments to observe how a full organ reacts to new antibodies and antibiotics, and reduce the time and risks put upon new drug developments. We see a better research pathway for regenerative medicine and stem cell technology as well as more research in general for future devices and techniques.

There are indeed possible challenges the current three-dimensional bioprinting must overcome. It will also encounter objections before becoming clinically applied because of legal and ethical ambiguities. Due to controversies in the application of the technology, government intervention may be needed to determine the status of the technology. With approval and consideration, clinical application is expected to spread rapidly in the service of humankind.[8,13] It will also be necessary to create national guidelines to regulate the technology and reduce as many ethical risks as possible. Through the application of three-dimensional bioprinting organs and their transplantation, we expect the life spans of patients with organ defects or dysfunctions to be elongated.[14] Yet, future research involving experiments, and application cases will be a crucial part of this new possible medical paradigm.

## Resources